# How do we drive deep eutectic systems towards an industrial reality?

Alexandre Paiva [1], Ana A. Matias [2,3], Ana Rita C. Duarte [1*]

[1] LAQV-REQUIMTE, Departamento de Química, Faculdade de Ciências e Tecnologia, Universidade Nova de Lisboa, 2829-516 Caparica, Portugal

[2] Instituto de Tecnologia Química e Biológica António Xavier, Universidade Nova de Lisboa, Av. da República, 2780-157 Oeiras, Portugal

[3] iBET, Instituto de Biologia Experimental e Tecnológica, Apartado 12, 2780-901 Oeiras, Portugal

*aduarte@fct.unl.pt

**ABSTRACT**

Deep eutectic systems (DES) have received considerable attention in the past 5 years of research. Up to 2013 little had been explored over these systems both in which concerns fundamentals and applications. The definition of a deep eutectic system is still controversial and in this manuscript we highlight the different definitions proposed in the literature and discuss what, in our opinion, should be pursued in order to have an agreement in the scientific community studying this field. From the definition of DES to its applications also interesting changes have been observed in the topics published. From electrochemistry to newer applications, such as cryopreservation, the field is evolving as new outcomes on DES properties are unraveled. We herein present and discuss new trends on DES research and present some perspectives on what we consider to be the most promising applications of DES at industrial scale.

# INTRODUCTION

Etymologically the word eutectic comes from greek eu - "ευ" and teksis - "Τήξις", meaning easy melting and it describes an homogeneous mixture of two solid species, which at a unique molar ratio form a joint super-lattice structure, which confers them peculiar properties. At this single point, the system melts as a whole and the solid systems become liquid. Figure 1 presents a general scheme of the DES preparation and the phase behavior in terms of melting point versus the ratio between the two constituents that form the system.

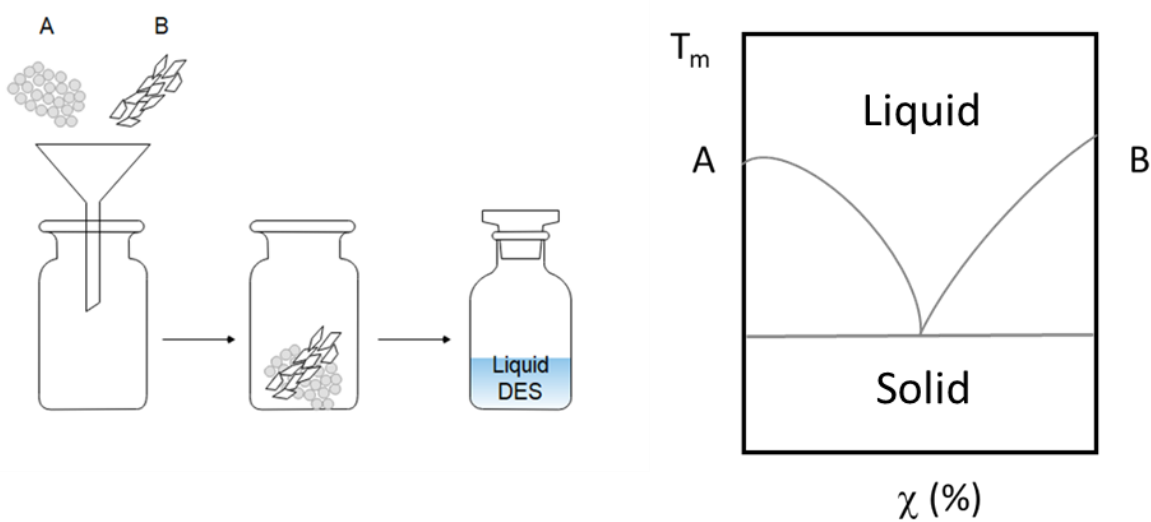

**Figure 1.** Schematic representation of the DES preparation and the phase behavior in terms of melting point versus the molar ratio of the two components

# THE DEFINITION OF A DES - THE RELEVANCE OF THE FUNDAMENTALS

DES have been gaining increased interest in the research community in the past 5 years[1]. However most of the publications are centered in the application of DES. Research dealing with the characterization of DES system interactions and properties is still scarce and even the definition of what is a DES in under dispute[2]. Two main denominations are used for these solvents; the most common deep eutectic solvent (DES) and another one proposed by Maria Francisco *et al.,* low transition temperature mixtures (LTTM)[3]. Although different denominations are used the principle applied to define these solvents is the same. A DES or LTTM is formed by two or more compounds that when mixed at a determined molar ratio suffer a high melting temperature depression. This decrease in the melting temperature of the mixture, compared to the individual components, is mainly attributed to hydrogen bond formation between the components.

The most commonly used DES are based on choline chloride which is highly hydroscopic. It is well known that water has a high influence on the hydrogen bond network of the DES and on the interactions between the components that form the DES. As a consequence, the presence of water in the system and its effect on the DES properties needs to subject to further studies.[4] Some authors have started to give some insights by nuclear magnetic resonance (NMR) and molecular dynamics in which concerns the behavior of DES at different conditions, namely at different molar ratios of the components or presence of water[5–7]. These authors confirm the importance of the hydrogen bond network in the formation of a DES, nevertheless what differs a DES from any eutectic mixture is still an unanswered question. So, the most important question that the scientific community most answer is in fact what is a DES? Figure 2 highlights the questions that in our opinion need to be addressed. Is it a mixture of two or more compounds that behave differently from an ideal mixture and suffer a depression on the eutectic point? Is it a mixture of two or more compounds that independently of behaving as an ideal mixture or not, present an eutectic point at a much lower temperature than the melting point of the individual components? And if so what is the value? Is it any mixture liquid at room or near room temperature?

Another intriguing question related the DES formed when one of the components is a liquid. For example, is a glycerol based DES, liquid at room temperature, a DES or a solution? Is a mixture of two liquids, which are able to form hydrogen bonds between them and also suffer a depression on the melting point, also a DES?

On another hand, and regarding the molar composition of a DES questions still remain to be answered. Is a DES a mixture with the molar composition of the eutectic point? Or would it also be any composition at which is the mixture is in the liquidus phase at a room or near room temperature? Is there a thermodynamic definition of DES that can be explained by the interactions of the components which behave differently from an ideal mixture and have a defined eutectic point lower than the ideal mixture? Or is there a broader definition of a DES which includes ideal and non ideal mixtures just as long as the mixture is in the liquidus phase at room or near room temperature?

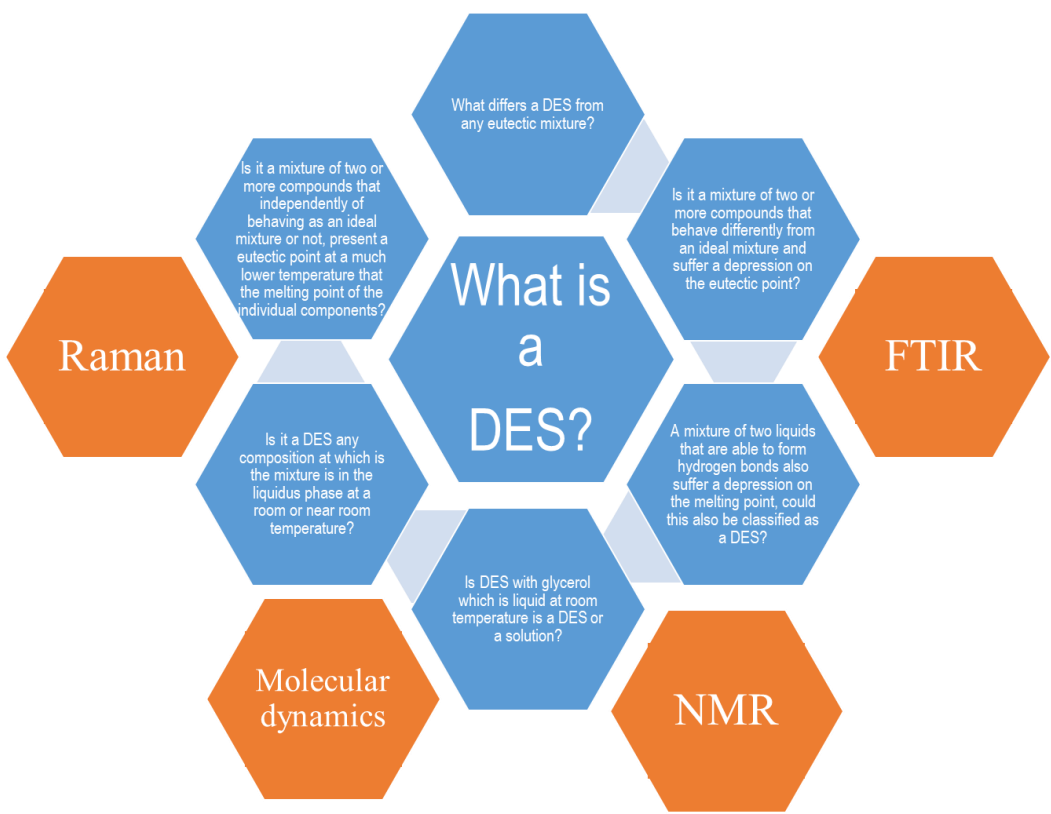

**Figure 2.** The definition of a DES: what remains unanswered and what techniques could provide valuable information on these systems

To answer these questions and to have a deeper understanding of the interaction between the components that form a DES, more fundamental research must be pursued by the scientific community. For example, a combination of NMR techniques such as pulsed-field gradient NMR, spin-relaxation times (T1) analysis or homonuclear Nuclear Overhauser (NOE) with Fourier-transform infra-red (FTIR) and RAMAN spectroscopy can be used to study intra and inter components interactions, molecular mobility and the effect of co-solvents such as water in the structure of the DES.

**WHAT DID WE LEARN – THE PAST 5 YEARS OF RESEARCH IN THE FILED**

The exploration of the potential of deep eutectic solvents and its applications has been fairly recent. Despite that, nearly 125 reviews have been written in the past five years. How much can we review from a topic that is so recent and has still so many questions to answer? A literature overview of the field can be summarized in figure 3. The number of publications per year has been steadily growing and the diversity of subjects being studied is also notable. This highlights the undoubtedly potential of the systems, but also how little do we know about them, how many questions we still need to answer and how much do we still need to learn.

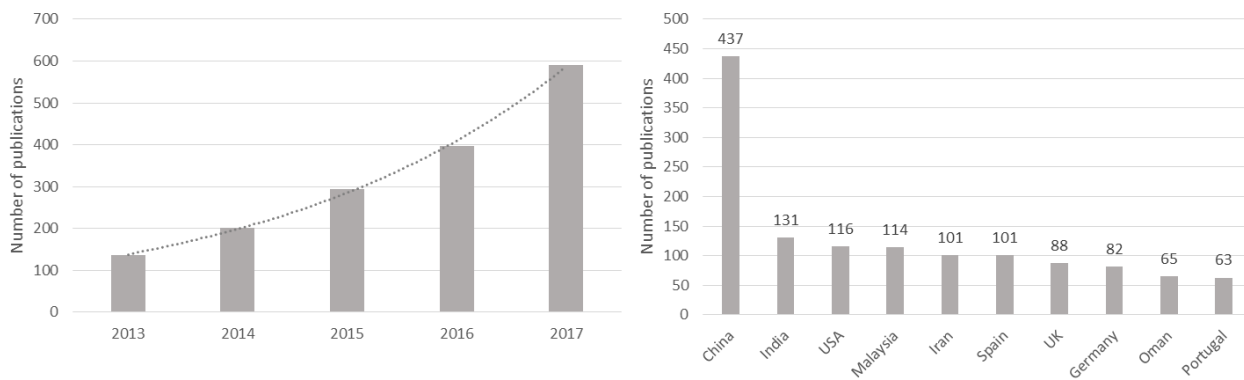

**Figure 3:** Overview of publications in "deep eutectic solvents": (left) number of publications per year (right) geographical distribution of the publications 2013-2017 (source: Web of Knowledge, December 2017)

Geographically, there is no doubt that Asia is by far ahead of Europe and USA. In terms of application development Asia is mostly devoting its resources towards energy fuels and electrochemistry and biotechnological applications, such as extraction of natural matrices or biomass processing, for example. A similar trend is observed in the research published in Europe. Europe has also published significantly in the fields of pharmaceutical sciences and biomedical applications. In the case of the USA, research has been more diverse and scattered and applied in less explored fields such as geochemistry, mining or mineralogy.

What we conclude from literature survey is that from the past five years of intense research in the field we have scattered funding and resources on application development rather than fundamental studies. The exponential growth on the number of publications reflects that this is an exciting and a largely unexplored field of research which may still render significant discoveries. However, it is our opinion that the use of DES should not proceed randomly as it has been the case. The scientific indicators from which researchers are evaluated is pushing the publication of scientific papers such that there is little time to answer fundamental questions. This is also the result of the lack of investment on fundamental science, which in turn leads to a poor understanding of the basic features of the systems, which is up to date still unknown. It is our perspective that in order to design the best systems for a given application it will be indubitably necessary a deeper knowledge on how do this systems form and how do they behave as well as the development of models able to simulate the systems and predict their properties.

## APPLICATIONS OF DEEP EUTECTIC SYSTEMS – WHAT COULD ACTUALLY BE SCALED UP?

The key advantages of DES over another alternative green solvents such ILs are that DESs have a lower cost and are much easier to produce in large scale batches, leveraging the scaling up of DES based processes in industrial environment[8,9]. Nevertheless, the scaling up of alternative greener processes are rather difficult taking many years to convince industry that they worthy, requiring, as any new developing technology, a number of steps between the initial concept and completion of the final production plant. The main operational steps include the scale-up from i) the bench to a pilot plant; ii) the pilot plant to the full scale process and optimization of the process. While the ultimate goal is to go directly from process optimization to full scale plant, the pilot plant is generally a necessary step. The information gathered from the pilot plant experiments allows a better overview of the overall process and this step helps to scale up the technology effectively and in a safely mode[8].

In particular, the application of DES is a rather young technology. Although significant progress has been made at lab scale, the implementation in the industry of new processes has not been established to date and few pilot scale studies have been undertaken or are already published.

Metal processing using DES-based processes have been in the past few years, the subject of a number of EU research projects being one of the fields with a straight focus on scale-up and commercialization towards the metal finishing and metal extraction industries. One of the major applications for DESs is their use as alternative media for metals that are traditionally difficult to plate or process, or involve environmentally hazardous processes. Comparatively to aqueous electrolytes, DES show higher solubility of metal salts and absence of water, and high conductivity compared to non-aqueous solvents. Currently, DESs have proved to be suitable solvents for the deposition of diverse metals in particular zinc, copper, nickel, silver, gold and tin and several joint projects between academia and large companies such as Rolls Royce are still ongoing focusing on scaling up and process optimization. Nevertheless, the deposition of aluminum is still at the lab scale and further studies addressing the increase in deposition rate must be carried out[8,9].

Another promising field of research receiving much interest involving DES, is the removal of sulfuric compounds from fuel oil. The separation of aromatic and aliphatic hydrocarbons having the same number of carbon atoms is also a challenging process especially at low aromatic concentration. In the past few years, the separation of aromatics using liquid-liquid extraction methods has been investigated at lab scale and both the distribution ratio and selectivity results have demonstrated the benefits of DESs in this separation operation. Due to the promising results, Ali and co-workers already conducted studies at pilot scale by for the validation of previous laboratory scale results. Despite the good results, in order to industrially implement the use of DES for liquid-liquid extraction processes and particularly in separation of aromatics from fuel, more pilot plant scale experiments must be conducted and the operational parameters should be optimized using gasoline fuel itself instead of a model mixture. This could constitute a very important step towards the implementation of this method at the industrial scale[10].

Biocatalysis is another filed where the application of DES is in expansion and potentially feasible to scale up to industrial scale in the next few years.[11]

DESs can work as solvents, co-solvents or extracting solvents in specific biocatalytic reactions.[12] Remarkably, as solvents, DESs can activate and stabilize the enzymes achieving highest efficiencies. Up to date, several enzymes such as lipases, proteases and epoxide hydrolases have shown great catalytic performance in DES, demonstrating their potential to replace ILs and/or organic solvents in biocatalytic reactions. Nevertheless, some questions remains to answer and an in-depth understanding on how DESs can activate and stabilize enzymes or how the biocatalysis products may actually be separated from DES will promote the application of DESs in biocatalysis to laboratory or industrial scale[8,13].

**CONCLUSIONS**

The field of deep eutectic systems is an exciting and challenging scientific field that in the past couple of years has open doors to numerous perspectives. New scientific developments are expected from the intense research on the topic but there is surely the need to systemize the efforts being made worldwide and bring the global efforts to a concrete reality. Being these systems as promising as they are described, in twenty years, society should benefit from the investment currently being made and sustainable processes like metal processing, liquid-liquid extraction and biocatalysis should become an industrial reality.

## Acknowledgments

The authors gratefully acknowledge the financial support of FCT through the project Des.zyme - Biocatalytic separation of enantiomers using Natural Deep Eutectic Solvents (PTDC/BBB-EBB/1676/2014). The research leading to these results has received funding from the European Union Horizon 2020 under grant agreement number ERC-2016-CoG 725034 (ERC Consolidator Grant Des.solve).